\documentclass[12pt]{article}

\usepackage{graphicx}
\begin{document}

\begin{center}
{\bf Magnetic black holes in AdS space with nonlinear electrodynamics, extended phase space thermodynamics and Joule–Thomson expansion } \\
\vspace{5mm} S. I. Kruglov
\footnote{E-mail: serguei.krouglov@utoronto.ca}
\underline{}
\vspace{3mm}

\textit{Department of Physics, University of Toronto, \\60 St. Georges St.,
Toronto, ON M5S 1A7, Canada\\
Department of Chemical and Physical Sciences, University of Toronto,\\
3359 Mississauga Road North, Mississauga, ON L5L 1C6, Canada\\
Canadian Quantum Research Center, 204-3002 32 Ave Vernon, BC V1T 2L7, Canada} \\
\vspace{5mm}
\end{center}
\begin{abstract}
Thermodynamics of magnetically charged black holes in Anti-de Sitter space in an extended phase space is studied. The cosmological constant is considered as a pressure and the black hole mass is treated as the chemical enthalpy. The black hole thermodynamics is similar to the Van der Waals liquid–gas thermodynamics. Quantities conjugated to the nonlinear electrodynamics parameter and a magnetic charge are obtained. The first law of thermodynamics and the generalized Smarr relation take place. We investigate critical behavior of black holes and Joule--Thomson expansion. The Gibbs free energy, the Joule--Thomson coefficient and the inversion temperature are calculated.
\end{abstract}

\vspace{3mm}
\vspace{3mm}

\vspace{3mm}

\section{Introduction}

Black holes behave as a thermodynamics system \cite{Bardeen,Jacobson,Padmanabhan} and the black hole area and surface gravity are connected with entropy and temperature, respectively \cite{Bekenstein,Hawking}. In Anti-de Sitter (AdS) space-time (the cosmological constant is negative) phase transitions in black holes occur \cite{Page}. Also, gravity with the negative cosmological constant is dual to the conformal field theory (CFT) named as AdS/CFT correspondence or gauge/gravity duality \cite{Maldacena,Witten,Witten1}. This is a holographic principle which connects two kinds of physical theories, gravity and conformal field theory. AdS/CFT correspondence allows to solve some quantum chromodynamics problems  \cite{Kovtun} and condensed matter problems \cite{Kovtun1,Hartnoll}. The cosmological constant in AdS-gravity is a thermodynamics pressure conjugated to a black hole volume. The black hole phase transitions in such approach were studied in \cite{Dolan,Kubiznak,Mann,Teo} and it was shown that the black hole thermodynamics is similar to liquid-gas thermodynamics.

In this paper we study thermodynamics of black holes in extended phase space where the coupling of nonlinear electrodynamics (NED) is considered as a thermodynamics quantity. NED models allow to smooth singularities of charges and to take into account quantum gravity corrections. The first NED model is Born--Infeld electrodynamics \cite{Born} which coupled to AdS-gravity was considered in \cite{Fernando,Dey,Cai,Fernando1,Myung,Banerjee,Miskovic}. It was demonstrated that black hole thermodynamics is similar to Van der Waals fluid thermodynamics.
Here, we study the Joule–Thomson expansion of NED-AdS black holes with heating and cooling regimes. Some aspects of black hole Joule--Thomson expansion were studied in \cite{Aydiner,Yaraie,Rizwan,Chabab,Mirza}.

In section 2 the metric function and corrections to the Reissner--Nordstr\"{o}m solution are obtained. The first law of black hole thermodynamics in the extended phase space is studied in section 3. The thermodynamic magnetic potential and the thermodynamic conjugate to the NED parameter are obtained. We show that the generalized Smarr relation holds. The critical specific volume, critical temperature and critical pressure are found in section 4. The Gibbs free energy is analysed. The Joule--Thomson adiabatic expansion of black holes is investigated in Section 5.  Section 6 is a conclusion. In Appendix we discussed functions used in NED.

The units with $c=\hbar=1$, $k_B=1$ are used.

\section{AdS black hole solution}

We will consider the action of NED coupled to general relativity with the negative cosmological constant $\Lambda$
\begin{equation}
I=\int d^{4}x\sqrt{-g}\left(\frac{R-2\Lambda}{16\pi G_N}+\mathcal{L}(\mathcal{F}) \right),
\label{1}
\end{equation}
where $G_N$ is the Newton constant, $\Lambda=-3/l^2$ and $l$ is the AdS radius. Here, we propose the NED Lagrangian
\begin{equation}
{\cal L}(\mathcal{F}) =-\frac{\sqrt{2{\cal F}}}{2\beta}\ln\left(1+\beta\sqrt{2{\cal F}}\right),
\label{2}
\end{equation}
with ${\cal F}=F^{\mu\nu}F_{\mu\nu}/4=(B^2-E^2)/2$, where $E$ and $B$ are the electric and magnetic induction fields, respectively. As $\beta\rightarrow 0$ in Eq. (2) we get the Maxwell Lagrangian ${\cal L}_M=-\mathcal{F}$. There is the description of various NED in Appendix.
Varying action (1) with respect to metric and 4-potential $A_\mu$ ($F_{\mu\nu}=\partial_\mu A_\nu-\partial_\nu A_\mu$) we obtain field equations
\begin{equation}
R_{\mu\nu}-\frac{1}{2}g_{\mu \nu}R+\Lambda g_{\mu \nu} =8\pi G_N T_{\mu \nu},
\label{3}
 \end{equation}
\begin{equation}
\partial _{\mu }\left( \sqrt{-g}\mathcal{L}_{\mathcal{F}}F^{\mu \nu}\right)=0,
\label{4}
\end{equation}
with $\mathcal{L}_{\mathcal{F}}=\partial \mathcal{L}( \mathcal{F})/\partial \mathcal{F}$.
The energy-momentum tensor of electromagnetic fields is
\begin{equation}
 T_{\mu\nu }=F_{\mu\rho }F_{\nu }^{~\rho }\mathcal{L}_{\mathcal{F}}+g_{\mu \nu }\mathcal{L}\left( \mathcal{F}\right).
\label{5}
\end{equation}
We will consider space-time with the spherical symmetry with the line element squered
\begin{equation}
ds^{2}=-f(r)dt^{2}+\frac{1}{f(r)}dr^{2}+r^{2}\left( d\theta
^{2}+\sin ^{2}(\theta) d\phi ^{2}\right).
\label{6}
\end{equation}
The field tensor $F_{\mu\nu}$ has the radial electric field $F_{01}=-F_{10}$ and the radial
magnetic field $F_{23}=-F_{32}=q\sin(\theta)$, where $q$ is the magnetic charge, $B=q/r^2$ is the magnetic field of the magnetic monopole. Thus, we treat the black hole as a magnetic monopole.  The metric function is given by \cite{Bronnikov}
\begin{equation}
f(r)=1-\frac{2m(r)G_N}{r},
\label{7}
\end{equation}
where the mass function being
\begin{equation}
m(r)=m_0+\int_{0}^{r}\rho (r)r^{2}dr.
\label{8}
\end{equation}
In Eq. (8) $m_0$ is the integration constant (the Schwarzschild mass) and $\rho$ is the energy density. It should be noted that electrically charged black holes possessing Maxwell's weak-field limit have singularities \cite{Bronnikov}.

Making use of Eq. (5) the magnetic energy density plus the vacuum energy density due to the negative cosmological constant is given by
\begin{equation}
\rho=\frac{q}{2\beta r^2}\ln\left(1+\frac{\beta q}{r^2}\right)-\frac{3}{2G_Nl^2}.
\label{9}
\end{equation}
By virtue of Eqs. (8) and (9) we obtain the mass function
\begin{equation}
m(r)=m_0+\frac{\pi q^{3/2}}{2\sqrt{\beta}}+\frac{q}{2\beta}\left[r\ln\left(1+\frac{\beta q}{r^2}\right)-2\sqrt{\beta q}\arctan\left(\frac{\sqrt{\beta q}}{r}\right)\right]-\frac{r^3}{2G_Nl^2}.
\label{10}
\end{equation}
The black hole magnetic mass is defined by
\begin{equation}
m_M=\int_0^\infty \frac{q}{2\beta}\ln\left(1+\frac{\beta q}{r^2}\right)dr=\frac{\pi q^{3/2}}{2\sqrt{\beta}}.
\label{11}
\end{equation}
In according with Eq. (11) the magnetic energy is finite but at $\beta=0$ it becomes infinite. Thus, the NED parameter $\beta$
smoothes singularities. Making use of Eqs. (7) and (10) one finds the metric function
\begin{equation}
f(r)=1-\frac{2MG_N}{r}-\frac{G_N q}{\beta}\ln\left(1+\frac{\beta q}{r^2}\right) +\frac{2q^{3/2}G_N}{\sqrt{\beta}r}\arctan\left(\frac{\sqrt{\beta q}}{r}\right)+\frac{r^2}{l^2},
\label{12}
\end{equation}
with the total mass $M=m_0+m_M$. As $\beta\rightarrow 0$ the metric function (12) becomes the metric function of Maxwell-AdS black holes
\[
f(r)=1-\frac{2MG_N}{r}+\frac{q^2G_N}{r^2}+\frac{r^2}{l^2}.
\]
Making use of Eq. (12) at $\Lambda=0$ ($l\rightarrow \infty$) and as $r\rightarrow \infty$, we obtain the metric function
\begin{equation}
f(r)=1-\frac{2MG_N}{r}+\frac{q^2G_N}{r^2}-\frac{q^3\beta G_N}{6r^4}+\mathcal{O}(r^{-6})~~~\mbox{as}~r\rightarrow \infty.
\label{13}
\end{equation}
Equation (13) shows that $M$ can be treated as the ADM mass. From Eq. (13) one can find corrections to the Reissner--Nordstr\"{o}m solution.
The plots of metric function (12) are depicted in Fig. 1 at $m_0=0$, $G_N=1$, $l=10$.
\begin{figure}[h]
\includegraphics {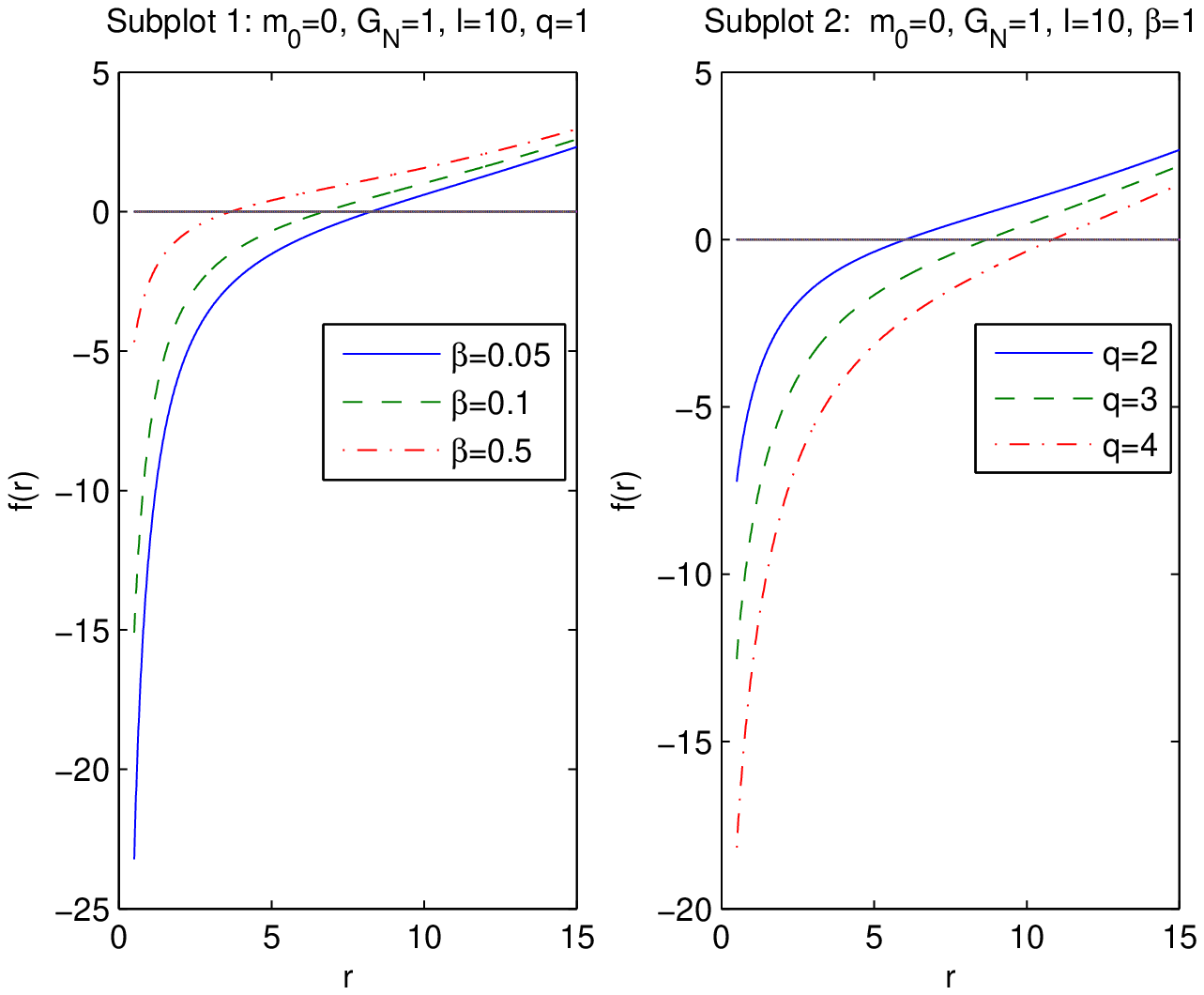}
\caption{\label{fig.1} The plots of the function $f(r)$ at $m_0=0$, $G_N=1$, $l=10$. Fig. 1 shows that there is only one black hole horizon. In accordance with subplot 1, when coupling $\beta$ increases the event horizon radius decreases. Subplot 2 shows that if magnetic charge $q$ increases the event horizon radius also increases.}
\end{figure}
According to Fig. 1 black holes possess one horizon radius $r_+$ ($f(r_+)=0$). When NED parameter $\beta$ increases (at constant $q$), the event horizon radius decreases. If magnetic charge $q$ increases (at constant $\beta$), the event horizon radius also increases.

\section{First law of black hole thermodynamics and the Smarr relation}

The generalized first law of black hole thermodynamics in extended phase space includes the pressure $P=-\Lambda/(8\pi)$, where $\Lambda$ is a
negative cosmological constant \cite{Kastor,Dolan1,Cvetic,Kubiznak1,Kubiznak2}, and is given by $dM=TdS+VdP+\Omega dJ+\Phi dq$ \cite{Kastor,Dolan1,Cvetic}. One has to interpret $M$ as a chemical enthalpy \cite{Kastor}, so that $M=U+PV$, where $U$ is the internal energy.
We obtain the Smarr relation from the first law of black hole thermodynamics exploring the Euler scaling argument \cite{Smarr} (see also \cite{Kastor}). Taking into account the dimensional analysis with $G_N=1$, we obtain $[M]=L$, $[S]=L^2$, $[P]=L^{-2}$, $[J]=L^2$, $[q]=L$, $[\beta]=L$. In extended phase space coupling $\beta$ is a thermodynamic variable. By using the Euler’s theorem \cite{Mann}, we find the mass
\begin{equation}
M=2S\frac{\partial M}{\partial S}-2P\frac{\partial M}{\partial P}+2J\frac{\partial M}{\partial J}+q\frac{\partial M}{\partial q}+\beta\frac{\partial M}{\partial \beta}.
\label{14}
\end{equation}
and $\partial M/\partial \beta\equiv {\cal B}$ is the thermodynamic conjugate to coupling $\beta$.  The black hole entropy $S$, volume $V$ and pressure $P$ are given by  \cite{Myers}, \cite{Myers1}
\begin{equation}
S=\pi r_+^2,~~~V=\frac{4}{3}\pi r_+^3,~~~P=-\frac{\Lambda}{8\pi}=\frac{3}{8\pi l^2}.
\label{15}
\end{equation}
By virtue of Eq. (12) we obtain the black hole mass
\begin{equation}
M(r_+)=\frac{r_+}{2G_N}+\frac{r_+^3}{2G_Nl^2}-\frac{qr_+}{2\beta}\ln\left(1+\frac{\beta q}{r_+^2}\right)+\frac{q^{3/2}}{\sqrt{\beta}}\arctan\left(\frac{\sqrt{\beta q}}{r_+}\right),
\label{16}
\end{equation}
where $r_+$ is the event horizon radius ($f(r_+)=0$).
When $\beta\rightarrow 0$ one finds the mass function of Maxwell-AdS magnetic black hole
\begin{equation}
M(r_+)=\frac{r_+}{2G_N}+\frac{r_+^3}{2G_Nl^2}+\frac{ q^2}{2r_+}~~~\mbox{as}~~\beta\rightarrow 0.
\label{17}
\end{equation}
Making use of Eq. (16) at $J=0$ for non-rotating black hole, we obtain (at $G_N=1$)
\[
dM(r_+)=\left[\frac{1}{2}+\frac{3r_+^2}{2l^2}-\frac{q}{2\beta}\ln\left(1+\frac{q\beta}{r_+^2}\right)\right]dr_+
-\frac{r_+^3}{l^3}dl
\]
\[
+\left[-\frac{r_+}{2\beta}\ln\left(1+\frac{q\beta}{r_+^2}\right)+\frac{3\sqrt{q}}{2\sqrt{\beta}}\arctan\left(\frac{\sqrt{q\beta}}
{r_+}\right)\right]dq
\]
\begin{equation}
+\left[\frac{qr_+}{2\beta^2}\ln\left(1+\frac{q\beta}{r_+^2}\right)-\frac{q^{3/2}}{2\beta^{3/2}}\arctan\left(\frac{\sqrt{q\beta}}{r_+}\right)\right]d\beta.
\label{18}
\end{equation}
The Hawking temperature is given by
\begin{equation}
T=\frac{f'(r)|_{r=r_+}}{4\pi},
\label{19}
\end{equation}
with $f'(r)=\partial f(r)/\partial r$.
From Eqs. (12), (19) and making use of equation $f(r_+)=0$ at $G_N=1$, we find the Hawking temperature
\begin{equation}
T=\frac{1}{4\pi}\biggl[\frac{1}{r_+}+\frac{3r_+}{l^2}-\frac{q}{\beta r_+}\ln\left(1+\frac{q\beta}{r_+^2}\right)\biggr].
\label{20}
\end{equation}
At the limit $\beta\rightarrow 0$ one has in Eq. (20) the Hawking temperature of Maxwell-AdS black hole. Making use of Eqs. (15), (18) and (20) one obtains the first law of black hole thermodynamics
\begin{equation}
dM = TdS + VdP + \Phi dq + {\cal B}d\beta.
\label{21}
\end{equation}
From Eq. (18) we obtain the magnetic potential $\Phi$ and the thermodynamic conjugate to the coupling $\beta$
\[
\Phi =-\frac{r_+}{2\beta}\ln\left(1+\frac{q\beta}{r_+^2}\right)+\frac{3\sqrt{q}}{2\sqrt{\beta}}\arctan\left(\frac{\sqrt{q\beta}}
{r_+}\right),
\]
\begin{equation}
{\cal B}=\frac{qr_+}{2\beta^2}\ln\left(1+\frac{q\beta}{r_+^2}\right)-\frac{q^{3/2}}{\beta^{3/2}}\arctan\left(\frac{\sqrt{q\beta}}{r_+}\right).
\label{22}
\end{equation}
At the limit $\beta\rightarrow 0$ in Eq. (22) one finds the magnetic potential of magnetic monopole $\Phi=q/r_+$.
The plots of potential $\Phi$ and ${\cal B}$ versus $r_+$ are depicted in Fig. 2.
\begin{figure}[h]
\includegraphics {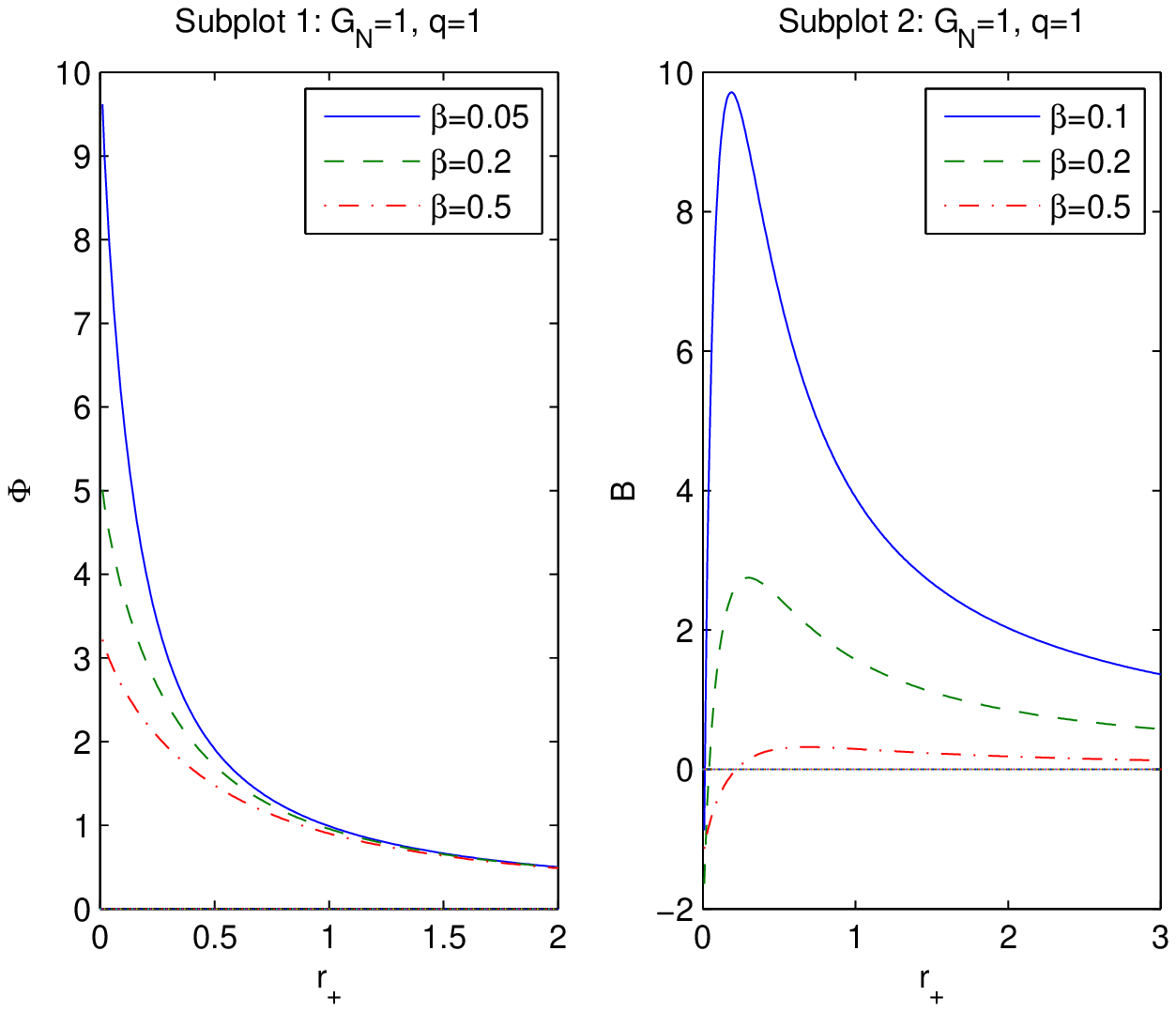}
\caption{\label{fig.2} The plots of the functions $\Phi$ and ${\cal B}$ vs. $r_+$ at $q=1$. In subplot 1, the solid curve is for $\beta=0.05$, the dashed curve corresponds to $\beta=0.2$, and the dashed-doted curve corresponds to $\beta=0.5$. The magnetic potential is finite at $r_+=0$ and vanishes at $r_+\rightarrow \infty$. When coupling $\beta$ increases the magnetic potential $\Phi$ decreases. In subplot 2, the function ${\cal B}$ becomes zero at $r_+\rightarrow \infty$ and finite at $r_+=0$. There are maxima of ${\cal B}$ at certain event horizon radii. }
\end{figure}
According to Fig. 2, subplot 1, when the coupling $\beta$ increases the magnetic potential $\Phi$ decreases. When $r_+\rightarrow \infty$ the magnetic potential $\Phi$ vanishes, $\Phi(\infty)=0$, and at $r_+ = 0$ it is finite.
In accordance with Fig. 2, subplot 2 when $r_+ = 0$ the vacuum polarization ${\cal B}$ is finite and as $r_+\rightarrow \infty$ it becomes zero, ${\cal B}(\infty)=0$. There are maxima of ${\cal B}$ at some event horizon radii.

Making use of Eqs. (15), (16) and (22), we obtain the generalized Smarr relation
\begin{equation}
M=2ST-2PV+q\Phi+\beta{\cal B}.
\label{23}
\end{equation}
Critical behavior of Born--Infeld-AdS black holes in extended phase space thermodynamics was studied in \cite{Mann1,Zou,Hendi,Hendi1,Zeng}.

\section{Black hole thermodynamics}

From Eq. (20) one finds the black hole equation of state (EoS)
\begin{equation}
P=\frac{T}{2r_+}-\frac{1}{8\pi r_+^2}+\frac{q}{8\pi\beta r_+^2}\ln\left(1+\frac{q\beta}{r_+^2}\right).
\label{24}
\end{equation}
At $\beta\rightarrow 0$ in Eq. (24), we obtain EoS for a charged  Maxwell-AdS black hole \cite{Kubiznak1}.
EoS (24) is similar to the Van der Waals EoS if we identify the specific volume $v$ with $2l_Pr_+$ \cite{Kubiznak1} ($l_P=\sqrt{G_N}=1$). Then Eq. (24) becomes
\begin{equation}
P=\frac{T}{v}-\frac{1}{2\pi v^2}+\frac{q}{2\pi\beta v^2}\ln\left(1+\frac{4q\beta}{v^2}\right).
\label{25}
\end{equation}
Equation (25) mimics the behaviour of the Van der Waals fluid. We find critical points, which are the inflection points in the $P-v$ diagrams, by equations
\[
\frac{\partial P}{\partial v}=-\frac{T}{v^2}+\frac{1}{\pi v^3}-\frac{q}{\pi\beta v^3}\ln\left(1+\frac{4q\beta}{v^2}\right)-\frac{4q^2}{\pi v^3(v^2+4 q\beta)}=0,
\]
\begin{equation}
\frac{\partial^2 P}{\partial v^2}=\frac{2T}{v^3}-\frac{3}{\pi v^4}+\frac{3q}{\pi\beta v^4}\ln\left(1+\frac{4q\beta}{v^2}\right)+\frac{8q^2}{\pi v^4(v^2+4 q\beta)}+
\frac{4q^2(12q\beta +5v^2)}{\pi v^4(v^2+4 q\beta)^2}=0.
\label{26}
\end{equation}
With the help of Eq. (26) we obtain the critical points equation
\begin{equation}
\frac{q}{\beta }\ln\left(1+\frac{4q\beta}{v_c^2}\right)+\frac{4q^2(5v_c^2+12q\beta)}{(v_c^2+4 q\beta)^2}-1=0.
\label{27}
\end{equation}
Making use of Eq. (26) one finds the equation for the critical temperature and pressure
\begin{equation}
T_c=\frac{1}{\pi v_c}-\frac{q}{\pi\beta v_c}\ln\left(1+\frac{4q\beta}{v_c^2}\right)-\frac{4q^2}{\pi v_c(v_c^2+4 q\beta)},
\label{28}
\end{equation}
\begin{equation}
P_c=\frac{1}{2\pi v_c^2}-\frac{q}{2\pi\beta v_c^2}\ln\left(1+\frac{4q\beta}{v_c^2}\right)-\frac{4q^2}{\pi v_c^2(v_c^2+4 q\beta)}.
\label{29}
\end{equation}
Solutions (approximate) to Eq. (27) for $v_c$, critical temperatures and pressures are presented in Table 1.
\begin{table}[ht]
\caption{Critical values of the specific volume and temperature at $q=1$}
\centering
\begin{tabular}{c c c c c c c }\\[1ex]
\hline
$\beta$ & 0.1 & 0.3 & 0.5 & 0.7 & 0.9 & 1 \\[0.5ex]
\hline
$v_{c}$ &4.848 & 4.743  & 4.637 & 4.528 & 4.416 & 4.359\\[0.5ex]
\hline
$T_{c}$ &0.0436 & 0.0442 & 0.0448 & 0.0454 & 0.0460 & 0.0464\\[0.5ex]
\hline
$P_{c}$ &0.0034  & 0.0035 & 0.0036 & 0.0037 & 0.0038 & 0.0039 \\[0.5ex]
\hline
\end{tabular}
\end{table}
The plots of $P-v$ diagrams are given in Figs. 3.
\begin{figure}[h]
\includegraphics {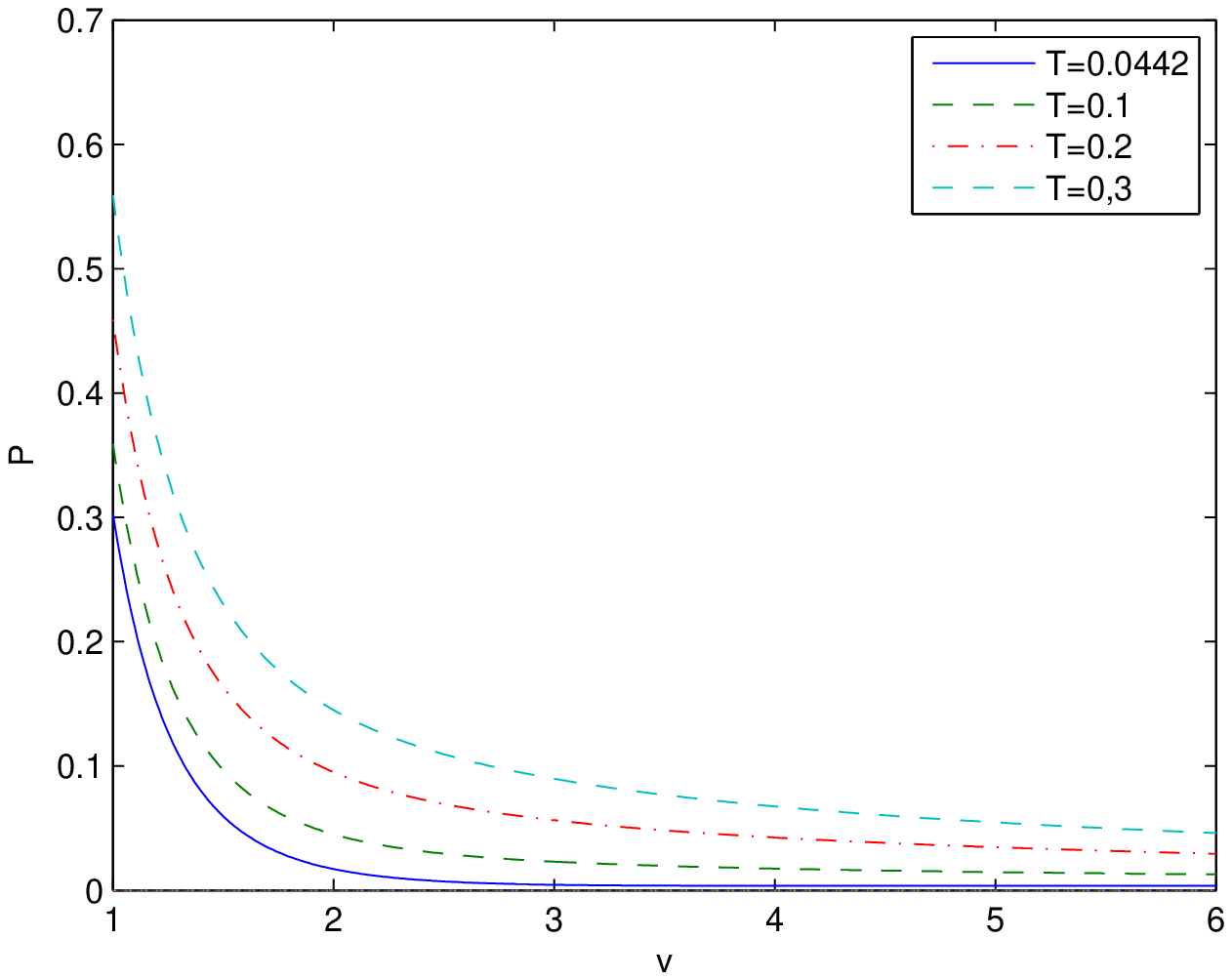}
\caption{\label{fig.3} The plots of the function $P$ vs. $v$ at $q=1$, $\beta=0.3$. The critical isotherm is for $T_{c}\approx 0.0442$ possessing the inflection point.}
\end{figure}
At $q=1$, $\beta=0.3$ the critical value for specific volume is $v_{c}\approx 4.743$ ($T_{c}=0.0442$).
By virtue of Eqs. (27), (28) and (29) and making use of Taylor series for small $\beta$ we find critical values of the specific volume, temperature and pressure
\begin{equation}
v_c^2=24q^2+{\cal O}(\beta),~~~T_c=\frac{1}{3\sqrt{6}\pi q}+{\cal O}(\beta),~~~
P_c=\frac{1}{96\pi q^2}+{\cal O}(\beta).
\label{30}
\end{equation}
At $\beta=0$ in Eq. (30) we find the critical points which are similar to charged AdS black hole points \cite{Mann1}. From Eq. (30) one obtains the critical ratio
\begin{equation}
\rho_c=\frac{P_cv_c}{T_c}=\frac{3}{8}+{\cal O}(\beta),
\label{31}
\end{equation}
with the value $\rho_c=3/8$ for the Van der Waals fluid.


The Gibbs free energy for a fixed charge, coupling $\beta$ and pressure ($M$ is a chemical enthalpy) is given by
\begin{equation}
G=M-TS.
\label{32}
\end{equation}
With the aid of Eqs. (16), (20) and (32) ($G_N=1$) we obtain
\begin{equation}
G=\frac{r_+}{4}-\frac{2\pi r_+^3P}{3}-\frac{qr_+}{4\beta}\ln\left(1+\frac{q\beta}{r_+^2}\right)+\frac{q^{3/2}}{\sqrt{\beta}}\arctan\left(\frac{\sqrt{q\beta}}{r_+}\right).
\label{33}
\end{equation}
At the limit $\beta\rightarrow 0$ Eq. (33) is converted to the Gibbs free energy of Maxwell-AdS black hole. The plots of $G$ versus $T$ with $\beta=0.3$ and $v_c\approx 4.743$, $T_c\approx 0.0442$ are depicted in Fig. 4. We took onto consideration, according to Eq. (24), that $r_+$ is a function of $P$ and $T$.
\begin{figure}[h]
\includegraphics {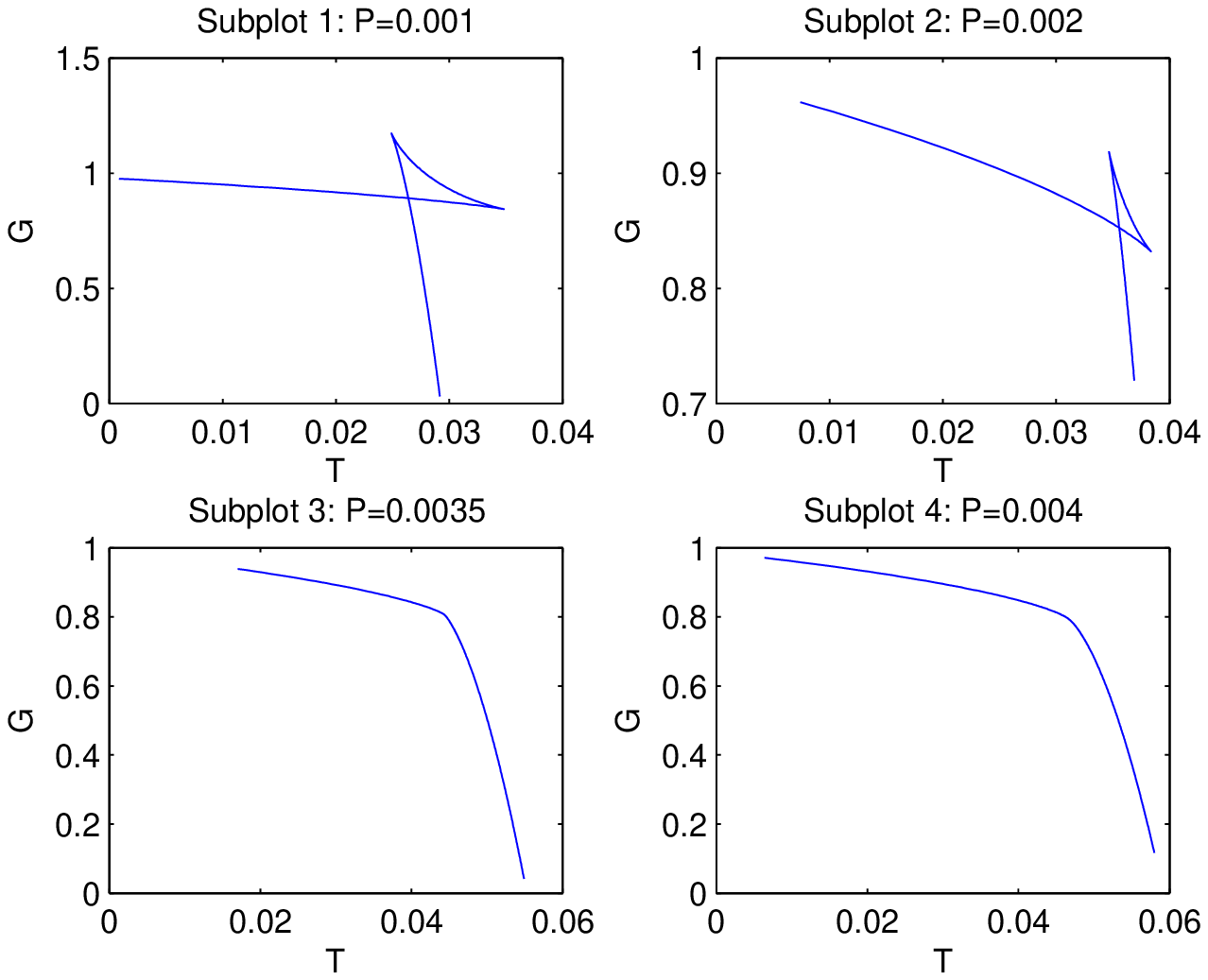}
\caption{\label{fig.4} The plots of the Gibbs free energy $G$ vs. $T$ with $q=1$, $\beta=0.3$. Subplots 1 and 2 show the critical 'swallowtail' behaviour with first-order phase transitions between small and large black holes. Subplot 3 corresponds to second-order phase transition ($P=P_c\approx 0.0035$) and subplot 4 corresponds to the case $P>P_c$ with non-critical behavior of the Gibbs free energy.}
\end{figure}
Subplots 1 and 2 at $P<P_c$ show first-order phase transitions, with 'swallowtail' behaviour, between small and large black holes. Subplot 3 is for $P=P_c$, where the second order phase transition occurs. Subplot 4 shows that in the case $P>P_c$ there are not phase transitions.

We depict the plots of entropy $S$ versus temperature $T$ at $q=\beta=1$ in Fig. 5. In accordance with Fig. 5, subplots 1 and 2, entropy is ambiguous function of the temperature in some intervals. Therefore, for this region first-order phase transitions take place.  Subplot 3 in Fig. 5 shows the second-order phase transition.  A low-entropy state and a high-entropy state are separated by the critical point. Figure 5, subplot 4, shows that there is not a critical behaviour of a black hole for these parameters, $q=\beta=1$, $P=0.006$.
\begin{figure}[h]
\includegraphics {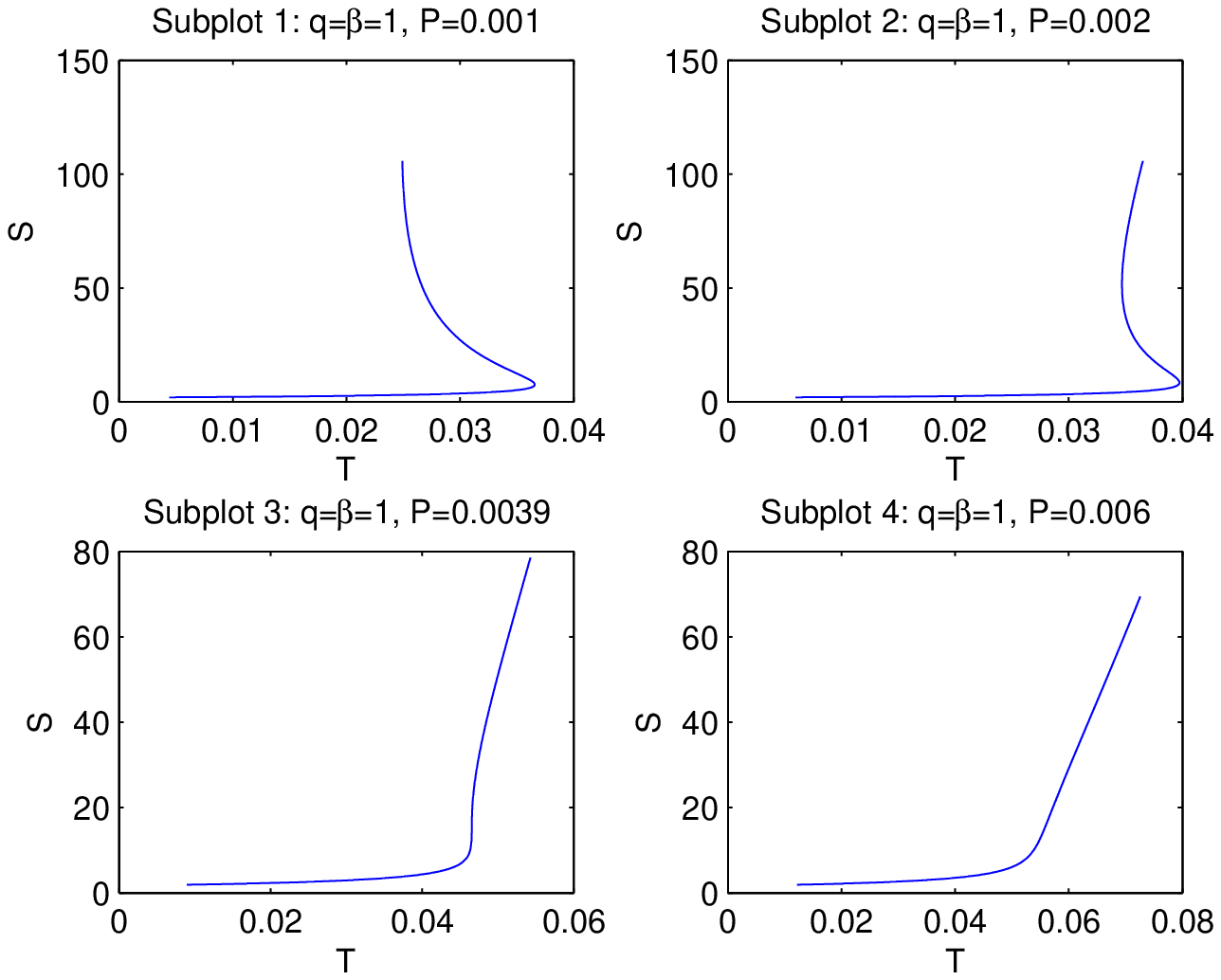}
\caption{\label{fig.5} The plots of entropy $S$ vs. temperature $T$ at $q=1$, $\beta=1$.}
\end{figure}

\section{Joule--Thomson expansion}

During the Joule--Thomson expansion the enthalpy which is the mass $M$, is constant. The Joule--Thomson thermodynamic coefficient characterises the cooling-heating phases and is given by
\begin{equation}
\mu_J=\left(\frac{\partial T}{\partial P}\right)_M=\frac{1}{C_P}\left[ T\left(\frac{\partial V}{\partial T}\right)_P-V\right]=\frac{(\partial T/\partial r_+)_M}{(\partial P/\partial r_+)_M}.
\label{34}
\end{equation}
Equation (34) shows that the Joule--Thomson coefficient is the slope in $P-T$ diagrams.
At the inversion temperature $T_i$ ($\mu_J(T_i)=0$) the sign of $\mu_J$ is changed. When the initial temperature is higher than inversion temperature $T_i$ during the expansion, the final temperature decreases that is the cooling phase ($\mu_J>0$). If the initial temperature is lower than $T_i$, the final temperature increases and corresponds to the heating phase ($\mu_J<0$). Making use of Eq. (34) and taking into account equation $\mu_J(T_i)=0$, one obtains the inversion temperature
\begin{equation}
T_i=V\left(\frac{\partial T}{\partial V}\right)_P=\frac{r_+}{3}\left(\frac{\partial T}{\partial r_+}\right)_P.
\label{35}
\end{equation}
The inversion temperature represents a borderline between cooling and heating process and the inversion temperature
line crosses points in maxima of $P-T$ diagrams \cite{Yaraie,Rizwan}. Equation (24) can be presented in the form
\begin{equation}
T=\frac{1}{4\pi r_+}+2P r_+-\frac{q}{4\pi r_+\beta}\ln\left(1+\frac{q\beta}{r_+^2}\right).
\label{36}
\end{equation}
At $\beta=0$ Eq. (36) is converted to EoS for Maxwell-AdS balack holes.
By using Eq. (16) ($G_N=1$) and $P=3/(8\pi l^2)$ we obtain
\begin{equation}
P=\frac{3}{4\pi r_+^3}\left[M-\frac{r_+}{2}+\frac{qr_+}{2\beta}\ln\left(1+\frac{q\beta}{r_+^2}\right)
-\frac{ q^{3/2}}{\sqrt{\beta}}\arctan\left(\frac{\sqrt{q\beta}}{r_+}\right)\right].
\label{37}
\end{equation}
Making use of Eqs (36) and (37) we depicted the $P-T$ isenthalpic diagrams in Fig. 6. Figure 6 shows that the inversion $P_i-T_i$ curve goes through maxima of isenthalpic diagrams.
 \begin{figure}[h]
\includegraphics {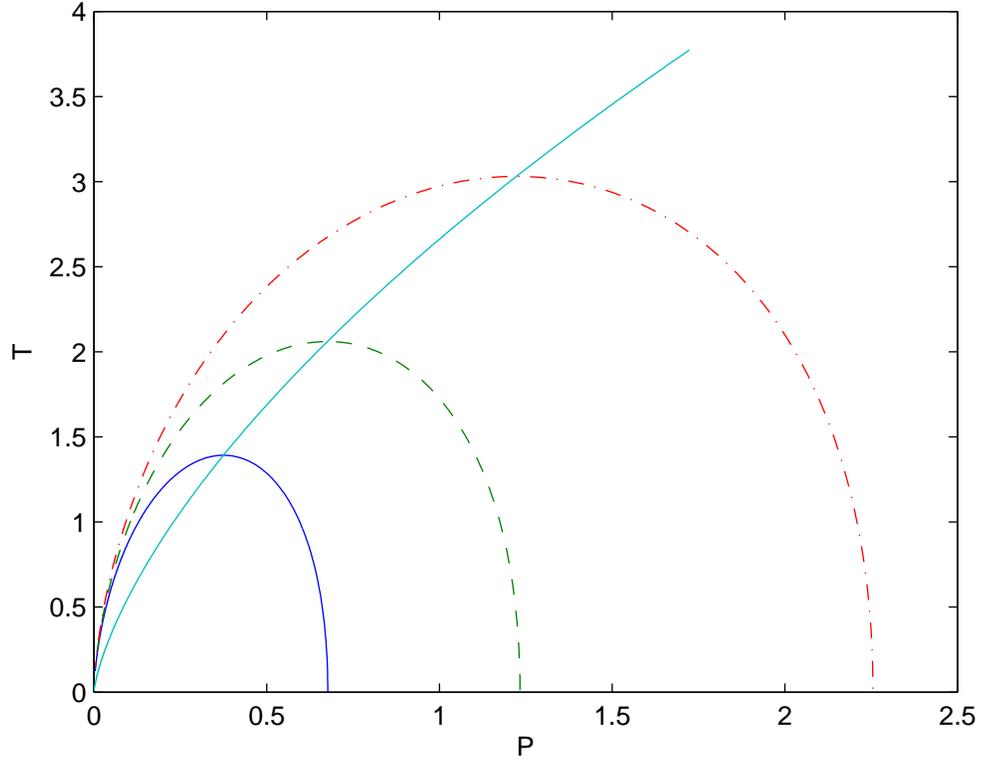}
\caption{\label{fig.6} The plots of the temperature $T$ vs. pressure $P$  and the inversion temperature $T_i$ with $q=20$, $\beta=0.5$.
The inversion $P_i-T_i$ curve goes through maxima of isenthalpic curves. The solid curve corresponds to mass $M=100$, the dashed curve
corresponds to $M=110$, and the dashed-doted curve corresponds to $M=120$. The plots of the inversion temperature $T_i$ vs. pressure $P_i$ for $q=20$, $\beta=0.5$ is presented by solid line.
When black hole masses increase the inversion temperature $T_i$ increases.}
\end{figure}
By virtue of Eqs. (24), (35) and (36) we obtain the equation for inversion pressure $P_i$
\begin{equation}
P_i=\frac{q}{4\pi\beta r_+^2}\ln\left(1+\frac{q\beta}{r_+^2}\right)+\frac{q^2}{8\pi r_+^2(r_+^2+q\beta)}-\frac{1}{4\pi r_+^2}.
\label{38}
\end{equation}
With the help of Eqs. (36) and (38) one finds the inversion temperature
\begin{equation}
T_i=\frac{q}{4\pi\beta r_+}\ln\left(1+\frac{q\beta}{r_+^2}\right)+\frac{q^2}{4\pi r_+(r_+^2+q\beta)}-\frac{1}{4\pi r_+}.
\label{39}
\end{equation}
Putting $P_i=0$ in Eq. (38), one finds the equation for the minimum of the event horizon radius $r_{min}$
\begin{equation}
\frac{q}{\beta}\ln\left(1+\frac{q\beta}{r_{min}^2}\right)+\frac{q^2}{2(r_{min}^2+q\beta)}-1=0.
\label{40}
\end{equation}
From Eqs. (39) and (40) at $\beta=0$ we obtain the inversion temperature minimum for Maxwell-AdS magnetic black holes
\begin{equation}
T_i^{min}=\frac{1}{6\sqrt{6}\pi q}, ~~~~r_+^{min}=\frac{\sqrt{6}q}{2}~~~~\mbox{at}~~\beta=0.
\label{41}
\end{equation}
At $\beta=0$ and from Eqs. (30) and (41), we obtain the relation $T_i^{min}=T_c/2$ which holds also for electrically charged AdS black holes \cite{Aydiner}.
Equations (38) and (39) represent parametric form of equations for $T_i$ versus $P_i$ which is given in Fig. 6. According to Fig. 6 the inversion point increases with increasing the black hole mass. The plots of the inversion curve $P_i-T_i$ with various parameters are depicted in  Figs. 7 and 8.
\begin{figure}[h]
\includegraphics {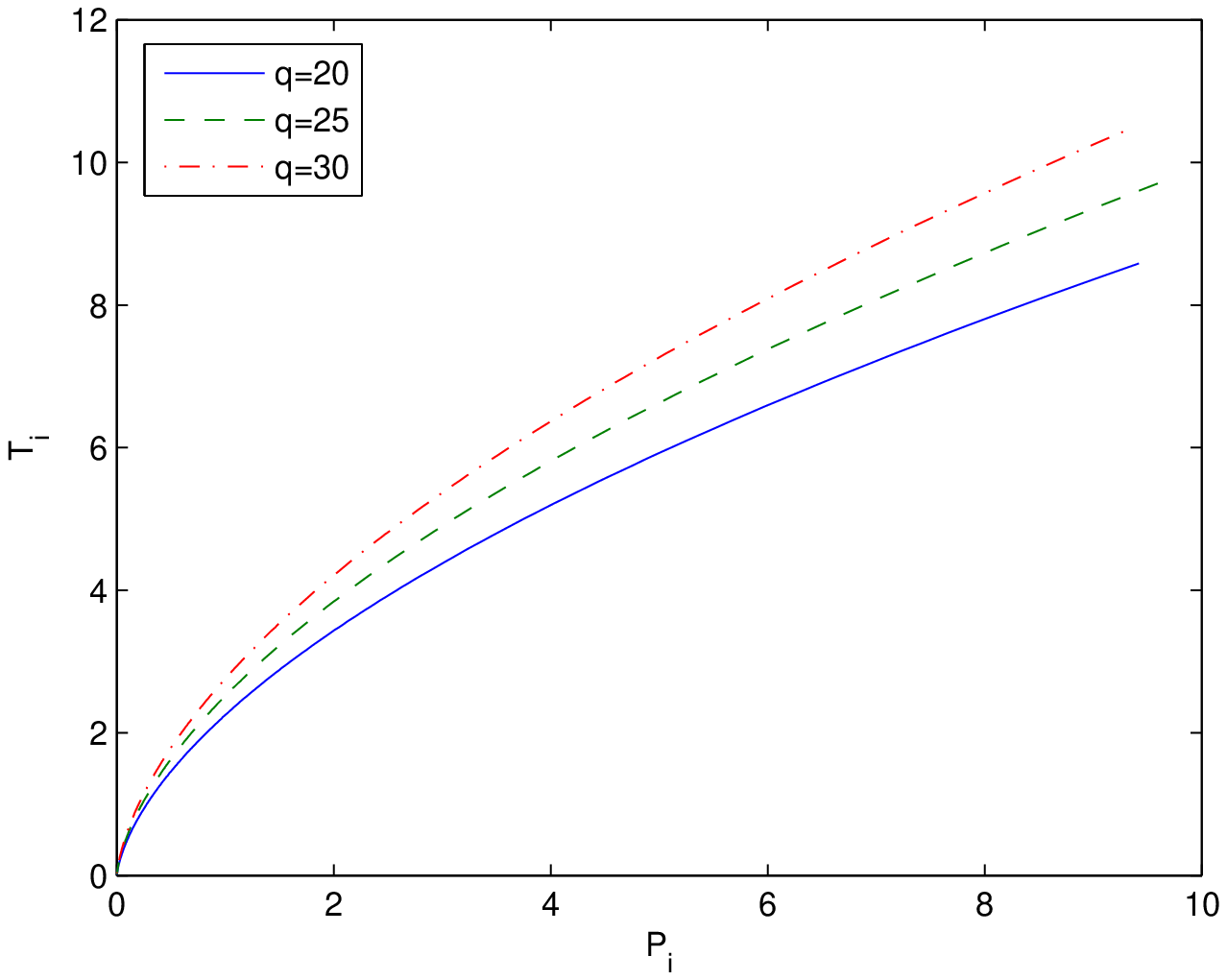}
\caption{\label{fig.7} The plots of the inversion temperature $T_i$ vs. pressure $P_i$ for $q=20$, $25$ and $30$, $\beta=1$. When magnetic charge $q$ increases the inversion temperature also increases.}
\end{figure}
\begin{figure}[h]
\includegraphics {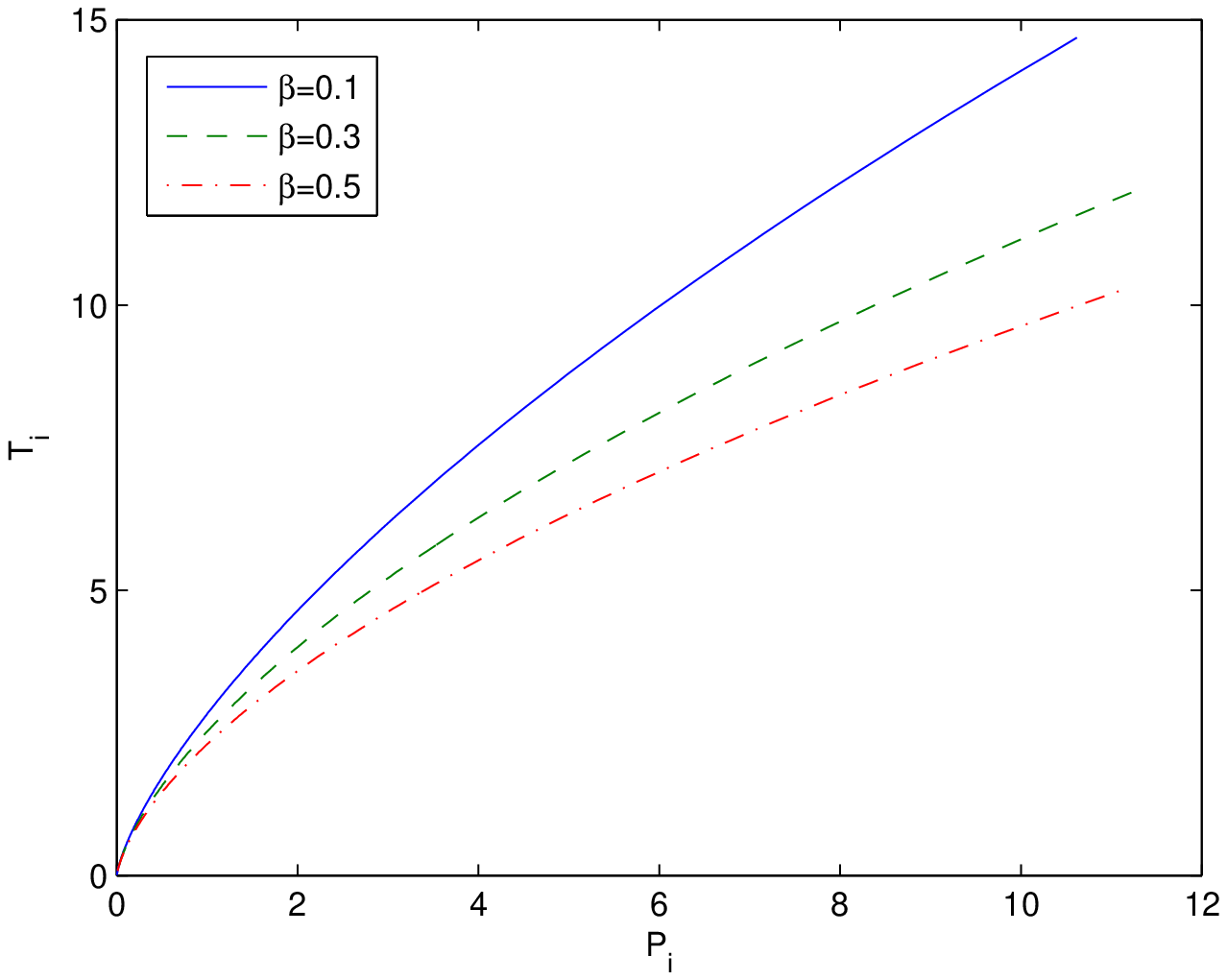}
\caption{\label{fig.8} The plots of the inversion temperature $T_i$ vs. pressure $P_i$ for $\beta=0.1$, $0.3$ and $0.5$, $q=15$. Figure shows that when the coupling $\beta$ increases then the inversion temperature decreases.}
\end{figure}
According to Fig. 7, with increasing magnetic charge $q$ (coupling $\beta$ is fixed) the inversion temperature also increases.
Figure 8 shows when the coupling $\beta$ increases at fixed magnetic charge $q$, the inversion temperature decreases.
From Eqs. (34), (36) and (37)  one finds the Joule--Thomson coefficient
\[
\mu_J=
\]
\begin{equation}
\frac{2r_+\left[\beta(r_+-6M)-qr_+\ln(1+q\beta/r_+^2)-6q^{3/2}\sqrt{\beta}\arctan (\sqrt{q\beta}/r_+)+\frac{q^2\beta r_+}{r_+^2+q\beta}\right]}
{3\left[\beta(r_+-3M)-qr_+\ln(1+q\beta/r_+^2)+3q^{3/2}\sqrt{\beta}\arctan (\sqrt{q\beta}/r_+)\right]}.
\label{42}
\end{equation}
If the Joule--Thomson coefficient is positive, $\mu_J>0$, a cooling process takes place. When $\mu_J<0$ a heating process occurs. In Fig. 6 the region with $\mu_J>0$ belongs to the left side of inversion temperature borderline and $\mu_J<0$ to the right site of borderline $T_i$.

\vspace{5mm}
\textbf{Appendix}
\vspace{5mm}

In this Appendix we present the short review of NED models with Lagrangians ${\cal L}(\mathcal{F})$. We only consider some models which give at weak-field limit Maxwell Lagrangian because Maxwell electrodynamics is well tested. Thus, we will discuss Lagrangians
\[
{\cal L}(\mathcal{F})=-\mathcal{F}+\beta\mathcal{F}^2~~~~\cite{Heisenberg,Kruglov5,Kr}~~~~(A1)
\]
\[
{\cal L}(\mathcal{F})= -\frac{1}{\lambda}\ln\left(1+\lambda\mathcal{F}\right) ~~~\cite{Soleng}~~~~~~~~~~~~~~~~~~~~~~~~~~~~~~~~~~~~~~~~~~~~~~~~~~~~(A2)
\]
\[
{\cal L}(\mathcal{F})=\frac{1}{2\beta}\biggl[\left(1-\sqrt{-2\beta\mathcal{F}}\right)\ln\left(1-\sqrt{-2\beta\mathcal{F}}\right)
\]
\[
+\left(1+\sqrt{-2\beta\mathcal{F}}\right)\ln\left(1+\sqrt{-2\beta\mathcal{F}}\right)\biggr]~~\cite{Gullu}~~~~~~~~~~~~~~~~~~~~~~~~~~~~~~~~~~~~~~~(A3)
\]
\[
{\cal L}(\mathcal{F})=-\frac{\mathcal{F}}{1+(2\beta\mathcal{F})^\gamma},~~~\gamma=1~~\cite{Kruglov},~~~\gamma=1/2,1/4~~\cite{Kruglov1}~~~~~~~~~~~~~~~~~~~~(A4)
\]
\[
{\cal L}(\mathcal{F})=
-\frac{\mathcal{F}}{(1+2\beta\mathcal{F})^\gamma},~\gamma=2~\cite{Kr6},~\gamma=3~\cite{Kr4},~\gamma=1/2~\cite{Kr7},~\gamma=3/2~\cite{Kr5}~(A5)
\]
\[
{\cal L}(\mathcal{F})=-\frac{1}{\beta}\arcsin(\beta\mathcal{F})~~~\cite{Kruglov2}~~~~~~~~~~~~~~~~~~~~~~~~~~~~~~~~~~~~~~~~~~~~~~~~~~~~~~~~~(A6)
\]
\[
{\cal L}(\mathcal{F})=-\frac{1}{\beta}\arctan(\beta\mathcal{F})~~~\cite{Kruglov3}~~~~~~~~~~~~~~~~~~~~~~~~~~~~~~~~~~~~~~~~~~~~~~~~~~~~~~~~(A7)
\]
\[
{\cal L}(\mathcal{F})=-\frac{\mathcal{F}}{\cosh^n\sqrt[4]{|\beta\mathcal{F}|}},~~~n=2 ~\cite{Bronnikov},~~n=1~\cite{Kr2}~~~~~~~~~~~~~~~~~~~~~~~(A8)
\]
\[
 {\cal L}(\mathcal{F})=\beta^2\left(\exp\left(-\frac{\mathcal{F}}{\beta^2}\right)-1\right)~~~\cite{Hendi}~~~~~~~~~~~~~~~~~~~~~~~~~~~~~~~~~~~~~~~~(A9)
 \]
 \[
 {\cal L}(\mathcal{F})=-\mathcal{F}\exp\left(-\beta\mathcal{F}\right)~~~\cite{Kr3}~~~~~~~~~~~~~~~~~~~~~~~~~~~~~~~~~~~~~~~~~~~~~~~~(A10)
 \]
\[
{\cal L}(\mathcal{F})= \frac{1}{\beta}\left[1-\left(1+\frac{\beta\mathcal{F}}{\sigma}
\right)^\sigma\right],~~\sigma=1/2~~\cite{Born},~~\mbox{arbitrary}~\sigma~~\cite{Kruglov4}~~~~(A11)
\]
Review of some NED Lagrangians was presented in \cite{Yang}. Lagrangian (A1) was used in \cite{Yajima,Ruffini,Magos} to construct static and spherically symmetric black hole solutions in the Einstein--Euler--Heisenberg system. This chose was used because NED (A1) appears in quantum electrodynamics. It was shown \cite{Soleng} that in the model (A2) the electromagnetic self-mass is finite. The dyonic solution in general relativity was obtained \cite{Kruglov8} in the framework of NED model (A2). It worth noting that modified logarithmic model proposed (2) is simpler as compared with \cite{Soleng}, generalized logarithmic model \cite{Kruglov9} and double-logarithmic model (A3). The mass and metric functions of NED coupled to Einstein gravity here are expressed in simple elementary functions. Rational electrodynamics (A4) ($\gamma=1$) explains the inflation of the universe \cite{Kruglov10} and gives the correct size of magnetic M87* black hole \cite{Kruglov11}. The NED model (A4) for $\gamma=1/2$ and $\gamma=1/4$ coupled to Einstein--Gauss--Bonnet gravity was studied in \cite{Kruglov12}, \cite{Kruglov13}. The Lagrangian (A5) was explored for an investigation of universe inflation and for description of magnetic black holes. The NED models (A5), (A6), A7) and (A8) lead to more complicated description of magnetic black holes compared to model (A4). Metric functions of exponential NED models (A9) and (A10) coupled to gravity contain special functions and also lead to complicated description of black holes. The interest to Born--Infeld-type electrodynamics (A11) is because it interpolates between Born--Infeld NED and exponential NED \cite{Kruglov14} but coupled to gravity possesses a complicated structure.

There are restrictions on couplings and fields associated with principles of causality and unitarity for $E=0$ \cite{Shabad} ${\cal L}_{\cal F}\leq 0$, ${\cal L}_{{\cal F}{\cal F}}\geq 0$, ${\cal L}_{\cal F}+2{\cal F}{\cal L}_{{\cal F}{\cal F}}\leq 0$. In the case $E=0$ the list of restrictions is as follows
\[
\beta\geq 0,~~~~\beta B^2\leq \frac{1}{3}~~~~\mbox{for}~~(A1),
\]
\[
\lambda\geq 0,~~~~\lambda B^2\leq 2~~~~\mbox{for}~~(A2),
\]
\[
\beta\geq 0,~~~~ B~\mbox{is~arbitrary}~~~~~~\mbox{for}~~(A3),
\]
\[
\beta\geq 0,~~\beta B^2\leq \frac{1}{3},~\gamma=1,~~ B~\mbox{is~arbitrary},~\gamma=1/2,1/4~~~\mbox{for}~~(A4),
\]
\[
\beta\geq 0,~\beta B^2\leq \frac{4-\sqrt{13}}{3}\approx 0.13,~\gamma=2,~\beta B^2\leq\frac{13-\sqrt{129}}{20}\approx 0.08,~\gamma=3,
\]
\[
\beta B^2\leq 2,~\gamma=1/2,~~\beta B^2\leq\frac{5.5-\sqrt{26.25}}{2}\approx 0.188,~\gamma=3/2~~\mbox{for}~~(A5),
\]
\[
\beta\geq 0,~~~~\beta B^2\leq 2~~~~\mbox{for}~~(A6),
\]
\[
\beta\geq 0,~~~~\beta B^2\leq \frac{2}{\sqrt{3}}~~~~\mbox{for}~~(A7),
\]
\[
\beta\geq 0,~~~~ B\leq \beta~~~~\mbox{for}~~(A9),
\]
\[
\beta\geq 0,~~~~\beta B^2\leq \frac{5-\sqrt{17}}{2}\approx 0.22~~~~\mbox{for}~~(A10),
\]
\[
\beta\geq 0,~~\sigma>0,~~~~~\frac{(1-2\sigma)\beta B^2}
{2\sigma}\leq 1~~~~\mbox{for}~~(A11).
\]
To have restrictions on $\beta B^2$ due to principles of causality and unitarity for model (A8), one needs numerical calculations.

\section{Summary}

We have found magnetic black hole solution in AdS space-time in the framework of new NED. Metric, mass functions and their asymptotic were obtained. It was demonstrated that the black hole has only one horizon with corrections to the Reissner--Nordstr\"{o}m solution. When coupling $\beta$ increases (at constant magnetic charge) the event horizon radius decreases, but if magnetic charge increases (at constant $\beta$) the event horizon radius increases. We studied black holes thermodynamics in an extended thermodynamic phase space where the cosmological constant is treated as a thermodynamic pressure and the mass of the black hole is the chemical enthalpy.  Thermodynamic quantity conjugated to coupling $\beta$ and thermodynamic potential conjugated to magnetic charge were obtained. We showed that the first law of black hole thermodynamics and the generalized Smarr formula take place. There is an analogy of black hole thermodynamics with the Van der Waals liquid–gas thermodynamics. The Gibbs free energy was evaluated and phase transitions were studied. We have calculated the critical ratio $\rho_c$ which is different from the Van der Waals value $3/8$. The black hole Joule--Thomson adiabatic expansion, cooling and heating phase transitions were investigated. The inversion temperature was found which separates cooling and heating processes of black holes during the Joule--Thomson expansion.

\end{document}